\documentclass[12pt]{article}
\usepackage{amssymb}
\usepackage{latexsym}
\usepackage{amsmath}
\usepackage[normalsize,hang,bf]{caption}
\usepackage{graphicx}

\newcommand{\SL}{\widetilde{{SL}(2,\mathbb{R})}}
\newcommand{\SLbis}{{SL}(2,\mathbb{R})}
\newcommand{\ls}{{\frak{sl}(2,\mathbb{R})}{}}

\newcommand{\h}{\,{\bf H}}
\newcommand{\x}{\,{\bf X}}
\newcommand{\y}{\,{\bf Y}}
\newcommand{\e}{\,{\bf E}}
\newcommand{\f}{\,{\bf F}}
\newcommand{\lb}{\,{\bf L}}

\renewcommand{\t}{\,{\bf T}}

\newcommand{\bz}{\,{\bf z}}
\newcommand{\be}{\,{\bf e}}
\newcommand{\ba}{\,{\bf a}}
\newcommand{\bn}{\,{\bf n}}
\newcommand{\bk}{\,{\bf k}}
\newcommand{\tr}{\mbox{\rm Tr\,}}
\newcommand{\ZZ}{\mathbb Z}
\newcommand{\RR}{\mathbb R}

\begin{document}

\begin{center} {\large Global geometry of the 2+1 rotating black hole \\}
\vspace{.5cm} P.~Bieliavsky${}^{a,}$\footnote{ E-mail :
pbiel@ulb.ac.be}, S.~Detournay${}^{b,}$\footnote{ E-mail :
Stephane.Detournay@umh.ac.be, "Chercheur FRIA" },
M.~Herquet${}^{b}$, M.~Rooman${}^{c,}$\footnote{ E-mail :
mrooman@ulb.ac.be, FNRS Research Director} and
Ph.~Spindel${}^{b,}$\footnote{ E-mail :
spindel@umh.ac.be}\\

\vspace{.3cm}  {${}^a${\it Service de G\'eom\'etrie diff\'erentielle}}\\
{\it Universit\'e
Libre de Bruxelles, Campus Plaine, C.P. 218}\\ {\it Boulevard du Triomphe,
B-1050 Bruxelles, Belgium}\\
\vspace{.2cm} {${}^b${\it
M\'ecanique et Gravitation}}\\ {\it Universit\'e de Mons-Hainaut, 20
Place du Parc}\\
{\it 7000 Mons, Belgium}\\
\vspace{.2cm}  {${}^c${\it Service de Physique th\'eorique}}\\
{\it Universit\'e
Libre de Bruxelles, Campus Plaine, C.P.225}\\ {\it Boulevard du Triomphe,
B-1050 Bruxelles, Belgium}\\
\end{center}
\vspace{.1cm}

\begin{abstract}

The generic rotating BTZ black hole, obtained by identifications
in $AdS_3$ space through a discrete subgroup of its isometry
group, is investigated within a Lie theoretical context. This
space is found to admit a foliation by two-dimensional leaves,
orbits of a two-parameter subgroup of $\SL$ and invariant under
the BTZ identification subgroup. A global expression for the
metric is derived, allowing a better understanding of the causal
structure of the black hole.

\end{abstract}

{\bf Introduction\\}

Vacuum Einstein equations in 2+1 dimensions with negative
cosmological constant admit black hole solutions \cite{DT},
referred to as BTZ black holes \cite{BTZ}. These solutions arise
from identifications of points of $AdS_3$ by a discrete subgroup
of its isometry group \cite{BTZ,BHTZ}. According to the type of
subgroup, the vacuum, extremal or generic (rotating or
non-rotating) massive black holes are obtained \cite{BHTZ}.

The non-rotating massive black hole was investigated in \cite{BRS}
in the framework of Lie groups and symmetric spaces. This approach
revealed a natural foliation (a trivial fibration) of $AdS_3$ by
2-dimensional leaves invariant under the action of the
identification subgroup and admitting a regular Poisson structure.
These surfaces were constructed as orbits of a bi-action of $\SL$
twisted by an external automorphism. As a result a global
expression of the metric of the black hole region between the
singularities was obtained, which yielded a dynamical picture of
the non-rotating black hole evolution.

In this paper, we apply a similar group theoretical approach to
generic rotating black holes. Here again, we obtain an intrinsic
foliation adapted to the residual isometry of the black hole
solution. However, the foliation is quite different than in the
non-rotating case. Indeed, the leaves are orbits of a
two-parameter subgroup of $\SL$. This foliation naturally leads to
a globally defined metric (up to polar angle singularities),
improving our understanding of the causal structure of such black
holes.

\vspace{.5cm}{\bf  BTZ-adapted foliation of $AdS_3$ \\}

Rotating BTZ black holes are obtained from identifications in
$AdS_3$ space. This space is isometric to $\SL$ endowed with its
invariant metric. This metric (the Killing metric), denoted
$\beta$, is defined at the
identity {\be} by \\
$\beta_{\be}(\x,\y)={\frac 1 2} \tr({\bf
XY})$, where $\x$, $\y$ are elements of $\ls$. In the following we
will take the elements $\h , \e $ and $\f$ as generators of $\ls$,
defined by the commutation relations~:
\begin{equation}
[\h ,\e]=2\e \quad , \quad [\h,\f]=-2\f \quad ,\quad [\e,\f]=\h
\qquad .
\end{equation}
We will also use the generator $\t = \e - \f$.

Rotating black holes are characterized by their mass $M$ and
angular momentum $J$, with $|J|<M$. The identifications defining
them are performed by means of the action of a discrete subgroup
of its isometry group \cite{BTZ,BHTZ}, whose component connected
to the identity is isomorphic to $\SL \times \SL/Z_2$. This
discrete subgroup, referred to as BTZ subgroup, is isomorphic to
$\ZZ$ and generated by $\exp{(2 \pi m \Xi)}$, where $\Xi$ is the
Killing vector field \cite{BRS}:
\begin{equation}\label{Xi}
\Xi= L_+ \overline{\h} -  L_- \underline{\h}  \qquad ;
\end{equation}
$\overline{\h}$ and $\underline{\h}$ denote respectively the
right- and  left-invariant vector fields on $\SL$ associated to
the element $\h$ of its Lie algebra. The parameters $L_\pm$ are
related to $M$ and $J$ by $L_\pm=\sqrt{M\pm J}$, hence clarifying
the geometrical meaning of $M$ and $J$. The BTZ black holes are
thus obtained from the identifications~:
\begin{equation}\label{BTZ}
 \bz \sim \exp (2 \pi \, m \, L_+ \, \h)\, \bz \,
\exp(2 \pi \,  m \, L_- \, \h) \qquad ,
\end{equation}
with $m \in \ZZ$.

It is well known that $\SL$ admits a globally defined
decomposition, referred to as Iwasawa decomposition, which allows
to represent each element $\bz \in \SL$ in a unique way as the
product~:
\begin{equation}
\bz=\ba \, \bn \, \bk \qquad ,
\end{equation}
with~:
\begin{eqnarray}
 \ba \in A &=& SO(1,1) = \{\exp (\phi \h) \, | \,\phi \in \mathbb{R} \}  \qquad , \\
 \bn \in N &=& \{ \exp(u\e)\, | \, u \in \mathbb{R} \}  \qquad ,\\
\bk \in K&=&\widetilde{SO(2)}= \{ \exp(\tau \t)\, | \,\tau \in
\mathbb{R} \} \qquad .
\end{eqnarray}
This decomposition is not adapted to the BTZ bi-action
(\ref{BTZ}). Instead we devise a modified, BTZ-adapted, Iwasawa
decomposition defining a global diffeomorphism on $\SL$. It reads,
for $L_+
> L_-$, i.e. $J>0$ ~:
\begin{equation}\label{MICS}
\bz= \ba^{L_+} \, \bn  \, \bk \, \ba^{L_-}
 \qquad  .
\end{equation}
For $J<0$, the decomposition to consider is $\bz= \ba^{L_-} \, \bn
\, \bk \, \ba^{L_+}$; we hereafter assume $J>0$.

The decomposition (\ref{MICS}) naturally induces a foliation of
$\SL$ whose 2-dimensional leaves ${\cal F}_\tau$, corresponding to
constant $\tau$ sections, are  stable with respect to the action
of the BTZ subgroup. These leaves are obtained as follows. We
introduce the action $\nu$ of the subgroup $R=AN$ on $\SL$ as~:
\begin{equation} \label{action}
\nu : R \times \SL \rightarrow \SL : (\ba\bn,\bz) \mapsto
\nu_{\ba\bn}(\bz)=\ba^{L_+} \, \bn  \, \bz \, \ba^{L_-} \qquad .
\end{equation}
The orbits of this action are obtained by acting with $\nu$ on a
given element of the subgroup $K$ transverse to the leaves :
\begin{equation} \label{fiber}
{\cal F}_\tau  = \{ \nu_{\ba \bn} (\exp (\tau \t) ) \}\quad ,
\quad \forall \tau \in \RR \quad.
\end{equation}
In particular, the subgroup $R$ constitutes the leaf at $\bk=\be$.

\vspace{.5cm}{\bf Modified Iwasawa coordinate system (MICS) \\}

The BTZ-adapted map (\ref{action}) defines a global coordinate
system $(\tau, u, \phi)$ by~:
\begin{equation}
 \Phi : \RR^3 \rightarrow \SL : (\tau ,u,\phi) \mapsto
 \nu_{\exp (\phi \h)\exp (u \e)} (\exp (\tau \t) \qquad ,
\end{equation}
referred hereafter as Modified Iwasawa Coordinate System (MICS).
In these coordinates, the $AdS_3$ metric reads as~:
\begin{eqnarray} \label{metric}
&&ds^2 = - d\tau^2 - du d\tau -  2 u L_+ d\tau d\phi -
L_- \sin(2\tau) du d\phi \nonumber \\
&& \qquad \qquad \qquad +2 \left[M + L_+ L_- \left( \cos(2 \tau) -
u \sin (2 \tau) \right) \right] d\phi^2 \qquad .
\end{eqnarray}
The identification (\ref{BTZ}), yielding BTZ black holes, is
simply~:
\begin{equation} \label{btzperiod}
 (\tau, u, \phi) \mapsto (\tau, u, \phi + 2 \pi m) \qquad , \qquad
m \in \ZZ \qquad ;
\end{equation}
hence, when restricting the $\phi$-coordinate to $0 \le \phi < 2
\pi$, eq.(\ref{metric}) becomes a global expression for the metric
for the rotating BTZ black hole. The relation between the MICS
coordinates and more usual coordinate systems can easily be
obtained at the level of $\SLbis$ by using an explicit $\SLbis$
matrix representation.

Note furthermore that the rotating BTZ black hole is
canonically endowed with a regular Poisson structure, whose
characteristic foliation coincides with the foliation induced by 
the action $\nu$ defined in eq. (\ref{action}). 
In $(\tau, u, \phi)$ coordinates, the normalized Poisson structure 
reads as~:
\begin{equation}
\{ \,\, , \, \} = \frac 2 {L_++L_- \cos(2 \tau)} 
\partial u \wedge \partial \phi \qquad .
\end{equation}

\vspace{.5cm}{\bf Causal structure \\}

The identifications
(\ref{btzperiod}) are known to induce acausal
 regions containing closed
time-like curves passing through every point. The BTZ subgroup generator $\Xi$
given by eq.(\ref{Xi}) is time-like in these regions,
whereas it is space-like in the
physical regions. The boundaries between physical and non-physical
regions are the BTZ singularities defined by:
\begin{equation}\label{singularity}
{\cal S} = \{ \bz \in AdS_3 \quad  \mbox{\rm such that} \quad
\beta_{\bz} (\Xi, \Xi) = 0 \}  \qquad .
\end{equation}
In terms of MICS, these singularities are given by the
$\phi$-invariant surfaces :
\begin{equation}\label{singularity1}
q(\tau,u) \equiv M + L_+ L_- \left[ \cos(2 \tau) -u \sin (2 \tau)
\right]=0 \qquad ;
\end{equation}
$q(\tau,u)>0$ corresponds to causally safe regions and
$q(\tau,u) \le 0$ to regions with closed causal curves.

Let us now express the BTZ horizons $\cal H$ in terms of MICS. For
that purpose, we first note that a lightray passing through the
point ${\bf z_0}$ can always be written as~:
\begin{equation}
 \bz (s) = {\bf z_0} \exp (s \lb), \quad s \in \RR \qquad ,
 \end{equation}
with $\lb$ a future pointing null vector obtained by rotation of
$\e$ around the $\t$-axis, expressed as
\begin{equation}
 \lb = Ad(\exp(\kappa \t)) \e \qquad , \qquad 0 \leq \kappa < \pi \qquad .
\end{equation}
This null vector can also be seen as resulting from another null
vector boosted by a Lorentz rotation with axis along $\h$. These
two remarks allow us to parametrize the light-rays passing through
a point $\bz = \ba^{L_+} \bn \bk \ba^{L_-}$ of $AdS_3$, with
direction specified by $\kappa$, as:
\begin{equation}
\ell_{\bz}^{\kappa}(s) = \ba^{L_+} \, \bn \, \bk \, \exp\left[s
Ad(\exp(\kappa \t)) \e\right] \, \ba^{L_-} \qquad .
\end{equation}
The future and past light-cones at point $\bz$ are defined as~:
\begin{equation}
{\cal
  C}_{\bz}^{\pm} = \{ \ell_{\bz}^\kappa(s) \, | \, 0 \leq \kappa <
  \pi, \quad s \in \RR^\pm \} \qquad .
\end{equation}

The position of $\bz$ with respect to the BTZ horizons can be
determined according to the number of directions $\kappa$ allowing
the light-rays to escape a given singularity, either in the past
or in the future. If this number is zero, $\bz$ is situated behind
the horizon, whereas if this number is infinite, $\bz$ is before
the horizon. By continuity, the horizons are defined as the set of
points $\bz$ for which a finite number of $\kappa$'s permit to
escape the singularity for infinite values of $s$, i.e. for which
the equation of the intersections ${\cal C}_{\bz}^{\pm} \cap \cal
S$:
\begin{equation}\label{hor1}
\beta_{\ell_{\bz}^\kappa(s)} (\Xi, \Xi) = 0 \qquad ,
\end{equation}
has no solution for finite $s$. In terms of MICS, this equation
reads as~:
\begin{equation}\label{hor2}
 2 L_+ L_- \,  \beta_{\be}(Ad(l_{\bz}^\kappa(s)) \h,\h) + M =0 \quad .
\end{equation}
It is a second order equation in $s$, invariant for the
substitution $\bz \mapsto \ba \bz$, whose coefficients depend on
the coordinates $u$ and $\tau$ of the starting point
$\bz(u,\tau,\phi)$ of the light-ray and on its direction $\kappa$.
Requiring that eq. (\ref{hor2}) has no solution for finite $s$,
neither in the past nor in the future, comes to impose the
coefficients of $s$ and $s^2$ to vanish. In this way we obtain the
equations of the inner and outer horizons $\cal H^-$ and $\cal
H^+$~:
\begin{eqnarray}\label{hor3}
{\cal H}_1^+ &: \tau=m \pi \qquad , \qquad {\cal H}_2^+ &: u=-\tan(\tau)\qquad ,\\
{\cal H}_1^- &: \tau=\frac \pi 2 + m \pi \qquad , \qquad {\cal H}_2^- &: u=\cot(\tau)
\qquad , \qquad \nonumber
\end{eqnarray}
where $m\in\ZZ$.

Strikingly, the horizons of the rotating BTZ black hole correspond
to the union of left classes of the subgroups $R=AN$ and
$\overline{R}=A \overline{N}$ 
($\overline{N} = \{ \exp(v\f)\, \,v \in \mathbb{R} \}$)
 of the form $\bz \, R$ and $\bz \,
\overline{R}$, with $\bz = \exp (m \frac{\pi}{2} \t), m \in \ZZ$.

This geometrical discussion is summarized in the Penrose diagram
depicted in Fig. 1. The diagram is obtained by considering a
constant $\phi$ section and making the coordinate transformation
$u=\tan(p)$, with $- \frac \pi 2 \le p < \frac \pi 2 $. On this
diagram, each point corresponds to a circle (the orbit of the
point under the action of the one parameter subgroup generated by
$\phi$ translation). The horizons ${\cal H}^\pm$ are depicted by
straight lines inclined at 45 degrees, crossing the $\tau$-axis at
$\tau=m \frac \pi 2$. The singularities, defined in
(\ref{singularity1}), are situated beyond the inner horizons
${\cal H}^-$.

The causal structure of the BTZ spaces is obtained by considering
the two fields of directions
\begin{equation}
\frac{d \tau}{d u} = \frac{-( L_+ L_- u \sin(2 \tau) +  2
q(\tau,u) ) \pm (L_+ + L_- \cos(2 \tau)) \sqrt{2 q(\tau,u)}} {2 (2
q(\tau,u) + L_+^2 u^2)},
\end{equation}
with $q(\tau,u)$, defined in eq.(\ref{singularity1}), positive in
the causally safe region. These direction fields result from a
projection on the tangent subspace, at each point of the $(\tau,\
p)$ coordinate plane, of the light-cone generators, parallelly to
the $\partial_{\phi}$ direction. Note that as the vector field
$\partial_\phi$ is not orthogonal to the $(\tau, p)$ coordinate
surface, they are not simply given by the intersections of the
light-cones with this plane.

We would like to emphasize that causal curves in the BTZ geometry
are projected on the $(\tau, p)$ coordinate plane onto curves that
never leave the projected cones. Hence, this Penrose diagram
provides the causal structure of rotating BTZ black holes.

Finally, let us note that the leaves of the foliation
(\ref{action}), the surfaces of constant $\tau$, are flat. To see
this, notice that the induced metric on the surface $\tau =
\tau_0$ reads as~:
\begin{equation} \label{Mis1}
ds^2 = \frac{d\zeta d\phi}{2 L_+} +  \zeta d\phi^2 \qquad ,
\end{equation}
with $\zeta = 2 q(\tau_0,u)$ ($\tau_0 \ne m \frac{\pi}{2})$. We
can further perform the change of coordinates $U=\frac{e^{-2 L_+
\phi}}{2 L_+}$ and $V=\frac{\zeta e^{2 L_+  \phi}}{2 L_+}$, in
terms of which the metric becomes~:
\begin{equation}
ds^2 = - dU dV, \qquad \quad  0<U<\infty \qquad, \qquad  -
\infty<V<\infty \qquad ,
\end{equation}
which corresponds to half a Minkowski space. The identifications
according to the bi-action (\ref{BTZ}) read in these coordinates
as~:
\begin{equation}
 (U,V) \mapsto (U e^{-4 \pi m L_+}, V e^{4 \pi m L_+}) \qquad ,
 \end{equation}
They yield closed timelike curves for $V<0$. Thus,
when performing the identifications, the $\tau =$ constant
surfaces exhibit a Misner-space causal structure \cite{Mis,HE} due
to the fact that they intersect the singularity.

\vspace{.5cm}{\bf Perspectives\\}

We showed in this paper that $AdS_3$ and rotating BTZ black holes
possess a privileged foliation, the surfaces of constant $\tau$,
on which the group $R = AN$ acts (see (\ref{action})). A
remarkable property of this group is the existence of a star
product \cite{PB} which differs from the usual Rieffel-Moyal
product \cite{MR}. The latter product is indeed adapted only to
spaces admitting an action of $\RR ^d$.
The physical implications, namely in the context of string theory,
of the existence of such a noncommutative structure in BTZ spaces
will be addressed in the near future.

     \begin{figure}[hbt]
     \begin{center}

     \resizebox{!}{0.9\hsize}{\includegraphics{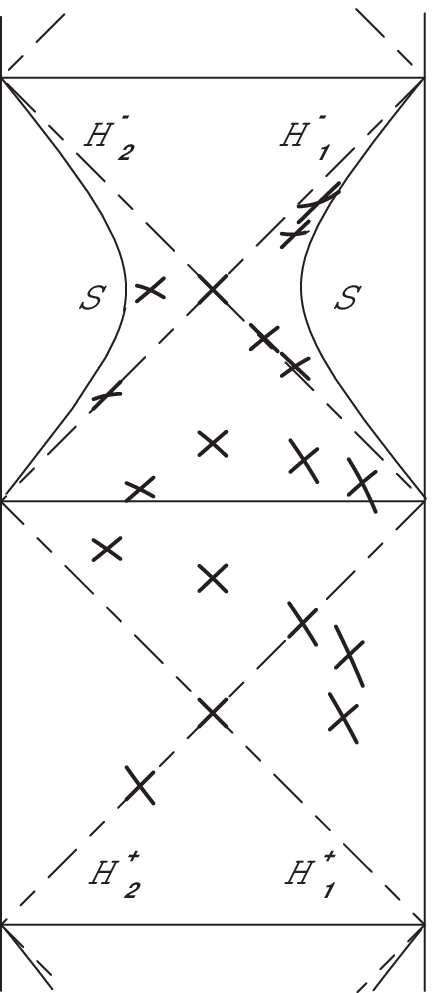}}
     \caption[]{\label{Penrose}The causal structure of the black hole is summarized
     in this Penrose diagram. The axis of the time-like coordinate
     $\tau$ is vertical, while that corresponding to the light-like coordinate
     $p$ is inclined at 45 degrees.
     Each point is a circle, orbit of the isometry subgroup generated by
     $\partial_{\phi} = L_+ \overline{\h} -  L_- \underline{\h}$.
       The projection parallel to $\partial_{\phi}$
     of the lightcone generators are depicted at several points.
     Curves always lying inside the projected light-cones can be
     lifted to causal curves joining a point ($\tau_0$,$p_0$,$\phi_0$)
     to a point($\tau_1$,$p_1$,$\phi_1$),
     with $\phi_0$ not necessarily equal to $\phi_1$.}

     \end{center}
     \end{figure}

\end{document}